\documentclass[%
 reprint,
 twocolumn,
nofootinbib,
 amsmath,amssymb,
 aps,
floatfix,
]{revtex4-1}

\usepackage{xspace}
\usepackage{xcolor}
\usepackage{caption}
\usepackage[normalem]{ulem}
\usepackage[breaklinks, plainpages=false, colorlinks=true, anchorcolor=cyan, linkcolor=magenta, citecolor=cyan, urlcolor=purple, bookmarks=false]{hyperref}
\usepackage{appendix}
\usepackage{graphicx}
\usepackage{acronym}
\usepackage{bm}
\usepackage{subcaption}
\usepackage{bbold}
\usepackage{tabularray}
\usepackage{soul}
\usepackage{lipsum}
\usepackage{physics}

\newacro{imbh}[IMBH]{intermediate-mass black hole}
\newacro{smbh}[SMBH]{supermassive black hole}
\newacro{bhns}[BHNS]{black hole neutron star}
\newacro{bbh}[BBH]{binary black hole}
\newacro{bh}[BH]{black hole}
\newacro{bns}[BNS]{binary neutron star}
\acrodef{FAR}[FAR]{false alarm rate}
\newacro{bf}[BF]{Bayes' factor}
\newacro{cbc}[CBC]{compact binary coalescence}
\newacro{ce}[CE]{Cosmic Explorer}
\acrodef{SNe}[SNe]{Supernova}
\newacro{da}[DA]{data analysis}
\newacro{et}[ET]{Einstein Telescope}
\newacro{eob}[EOB]{Effective-One-Body}
\newacro{fd}[FD]{frequency domain}
\newacro{gw}[GW]{gravitational wave}
\newacro{gr}[GR]{General relativity}
\newacro{hm}[HM]{Higher mode}
\newacro{ifo}[IFO]{interferometer}
\newacro{imr}[IMR]{inspiral-merger-ringdown}
\newacro{im}[IM]{inspiral-to-merger}
\newacro{kagra}[KAGRA]{Kamioka Gravitational Wave Detector}
\newacro{ligo}[LIGO]{Laser Interferometer Gravitational-Wave Observatory}
\newacro{lso}[LSO]{Last Stable Orbit}
\newacro{lvc}[LVC]{LIGO-Virgo Collaboration}
\newacro{lvk}[LVK]{LIGO-Virgo-KAGRA Collaboration}
\newacro{lo}[LO]{leading order}
\newacro{ns}[NS]{neutron star}
\newacro{nr}[NR]{numerical relativity}
\newacro{pn}[PN]{post-Newtonian}
\newacro{pe}[PE]{parameter estimation}
\newacro{psd}[PSD]{power spectral density}
\newacro{cwb}[cWB]{coherent waveburst}
\newacro{far}[FAR]{false alarm rate}
\newacro{ifar}[iFAR]{inverse false alarm rate}
\newacro{ml}[ML]{machine learning}
\newacro{cnn}[CNN]{convolutional neural network}
\newacro{asd}[ASD]{amplitude spectral density}
\acrodef{KN}[KN]{kilonova}
\newacro{xg}[XG]{next-generation}
\newacro{jsd}[JSD]{jensen shannon divergence}
\newacro{qnm}[QNM]{quasi-normal mode}
\newacro{cwt}[CWT]{continuous wavelet transform}
\newacro{hlv}[HLV]{Hanford-Livingston-Virgo}
\newacro{ecmm}[ECMM]{effective chirp mass model}
\acrodefplural{KN}[KNe]{kilonovae}
\newacro{qc}[QC]{quasi-circular}
\newacro{snr}[SNR]{signal-to-noise ratio}
\acrodef{SNR}[SNR]{signal-to-noise ratio}
\newacro{ng}[NG]{Next Generation}
\newacro{eos}[EoS]{Equation of State}
\newacro{agn}[AGN]{Active Galactic Nuclei}

\begin{document}

\title{Binary Black Hole inspirals can't hide their eccentricity} 

\author{Johann Fernandes}
    \email{johann.fernandes@iitb.ac.in}
    \affiliation{Department of Physics, Indian Institute of Technology, Bombay, Powai, 400076, India}
    \author{Praveer Tiwari}
        \email{ptiwari3009@gmail.com}
    \affiliation{Department of Physics, Indian Institute of Technology, Bombay, Powai, 400076, India}
\author{Archana Pai}%
    \email{archanap@iitb.ac.in}
    \affiliation{Department of Physics, Indian Institute of Technology, Bombay, Powai, 400076, India}


\begin{abstract}
   The events detected by the LIGO-Virgo-KAGRA collaboration over a period of 10 years have yielded a treasure trove of signals from compact binary coalescences. None of these events have shown a confident signature of eccentricity. With upgrades to the existing network and potential next-generation gravitational wave detectors, we will be able to see much further into the universe increasing the likelihood of detecting eccentric systems. We improve upon the phenomenological approach of providing eccentricity constraints using an effective chirp mass model in the time frequency domain. We introduce an improved pixel collection method along with a likelihood-based sampling approach inspired by Bayesian parameter estimation. Our approach constructs a likelihood from the product of energies collected across different eccentric harmonics in the time–frequency representation. This formulation enables coarse but meaningful constraints on orbital eccentricity. Additionally, we incorporate information from the energy ratios between eccentric harmonics, further refining the eccentricity estimates. We test our approach on 500 non-spinning equal mass eccentric systems and demonstrate that we can constrain the eccentricity within 0.2 around the true value. Moreover, our approach can deliver these constraints in 5 minutes on a machine with 50 cores. These results demonstrate that our phenomenological approach provides fast and reasonably accurate eccentricity estimates, making it a promising tool for rapid gravitational-wave data analysis.
\end{abstract}
\maketitle

\section{Introduction}
The \ac{lvk} has detected about 400 events in the \ac{gw} window over the course of its lifetime, spanning four observing runs \cite{gracedb, LIGOScientific:2014pky, Capote:2024rmo, VIRGO:2014yos, KAGRA:2020tym}. Most of the detected events are \acp{bbh} with a few \ac{bhns} and \ac{bns} systems. Parameter estimation studies carried out under the quasi-circular assumption of the binary provide an estimate of the astrophysical parameters \cite{LIGOScientific:2025slb, KAGRA:2021vkt}. While none of the events show a clear signature of eccentricity, some studies with numerical relativity simulations \cite{Gayathri:2020coq, Gamba:2021gap, Romero-Shaw:2021ual} suggest that GW190521 and GW190620 exhibit non-zero eccentricity. However, with the inclusion of higher order modes, this evidence of eccentricity vanishes \cite{Iglesias:2022xfc}.

The presence of eccentricity places strong constraints on the \acp{bbh} formation channel of which there are two major types, namely the isolated and dynamical formation channels. In the isolated formation channel, two stars in a binary system evolve to form a \ac{bbh} system that merges in Hubble time \cite{Peters:1963ux,Mapelli:2021taw}. The fate of these binaries is determined by the interactions between their individual components, leading to processes such as mass transfer or a common envelope phase. The resulting \ac{bbh} formed in such a scenario tends to display mass symmetry, aligned spins and a quasicircular orbit as any eccentricity is lost either by friction in the common envelope phase or by mass transfer processes \cite{hamers2019analytic, Sepinsky:2007in}. Dynamical formation channels, on the other hand, form binaries that have been significantly affected by collisional dynamics in dense stellar environments. Globular and nuclear star clusters are some examples of such environments that are dominated by strong gravitational interactions. \acp{bbh} formed in such environments are expected to display isotropic spin distributions \cite{Fishbach:2017dwv, Rodriguez:2016vmx}. Thus, in principle the component spins of the \acp{bbh} can hint at the formation channel. This is, however, not straightforward, as there can be overlap between the spin distributions of two channels, muddying conclusions. This is clear from the GWTC-4 population studies, which depict an asymmetry in the effective spin parameter, suggesting the presence of a preferentially aligned subpopulation \cite{LIGOScientific:2025pvj}. For instance, in isolated channels, if one of the components undergoes a supernova explosion imparting it a kick, it can lead to spin misalignment \cite{Gerosa:2018wbw}. Additionally, \cite{Stegmann:2020kgb} have proposed a model where mass transfers too can lead to misaligned spins. As a result spins alone cannot break the degeneracy between isolated and dynamical evolution of binary systems.

In contrast, eccentricity is a {\it tell-tale} signature of a dynamical formation channel \cite{Zevin:2021rtf}, making its confident detection an important milestone in the field of \ac{gw} astronomy. Despite this, searching for eccentric systems remains challenging as it adds three additional physical parameters; namely, initial eccentricity, the mean anomaly and the argument of periapsis \cite{Wagner:2024ecj}. Due to this, modelled searches that rely on template banks become computationally expensive as the number of templates required to cover the space increases dramatically \cite{Phukon:2024amh}. In contrast, model agnostic searches that use coherent excesses in energy across detectors better serve the problem at hand as they do not rely on template banks. These searches, when applied to LIGO data so far, did not recover any new significant candidates displaying eccentricity \cite{LIGOScientific:2019dag, LIGOScientific:2023lpe}. They were however able to place upper limits on the merger rate density of eccentric binaries from \ac{agn} and dense stellar cluster assisted mergers. Dealing with eccentricity is challenging on the parameter estimation front as well since the additional free parameters increase the cost of parameter estimation.

Multiple waveform models have been developed that include the effects of eccentricity \cite{Tanay:2016zog, Albanesi:2025txj, Nagar:2024oyk, Paul:2024ujx, Ramos-Buades:2021adz}. In this paper, we use TEOBResumS-Dali \cite{Albanesi:2025txj, Nagar:2024oyk} and EccentricTD \cite{Tanay:2016zog} for performing an injection study. These waveforms use different approaches to modelling the eccentric binary system. EccentricTD is a time-domain inspiral-only model supporting non-spinning systems up to an eccentricity of 0.9. It utilises a Keplerian-type parameterisation and modifies the existing TaylorT4 approximant to accommodate orbital eccentricities upto a 2PN order. It can accurately produce waveforms upto initial eccentricities of 0.9. On the other hand, TEOBResumS uses an \ac{eob} parameterisation to obtain an effective one-body Hamiltonian that describes the two-body problem. It is able to support a wide variety of systems, including non-circular, hyperbolic and non-planar orbits. TEOBResumS maintains accuracy in the late inspiral via the use of Pade resummations and a quasicircular \ac{nr} informed ringdown model.

In this work, we improve upon the phenomenological work proposed in \cite{Hegde2024, Bose2021} for constraining the eccentricity of stellar mass black hole binaries with moderate eccentricity. The authors had constructed a phenomenological model for the fundamental track and a scale factor to collect the energy along the eccentric harmonics in the time-frequency representation. In this work we improve upon this approach in two folds. First, we improve the pixel extraction method while collecting energy along with a likelihood-based sampling approach. Secondly, we fold the additional information of the energy ratio between the fundamental and the higher eccentric mode to further constrain the eccentricity. This method can act as a quick and alternative approach to full Bayesian parameter estimation in estimating eccentricity.

The paper is organised as follows. In section \ref{sec:theory}, we discuss the theory of eccentric inspiralling binaries and the systems we target in this work. Section \ref{sec:methodology} reviews existing phenomenological approaches to constraining eccentricity and provides a description of our model. We finally present the eccentricity constraints from our model in section \ref{sec:results}. We show that we can provide tight constraints on the eccentricity for nonspinning systems with a median eccentricity uncertainty of $0.2$. As a primer for future work in section \ref{sec:analytical}, we present an approach to deal with asymmetric systems, something that this current work does not take into account.

\section{Inspiraling Eccentric systems} \label{sec:theory}
\subsection{Theory of harmonic decomposition}
In this section, we briefly recapitulate the \ac{gw} emission from an eccentric binary system, restricting ourselves to the leading order calculation to develop an intuition of the signal morphology in the time-frequency representation, which we use in this work. 

Consider a binary system of two non-spinning black holes with masses $m_1$ and $m_2$ at a separation of $r$. Let $\phi, e$ and $a$ denote the true anomaly, eccentricity and semi-major axis of the binary orbit at any given instant, and
$i$ and $\Phi$ denote the inclination and azimuthal angles of the binary relative to the observer. Then, using the quadrupole formula in GW emission as shown in \cite{Moreno-Garrido1995}, one can obtain the two polarisations of GW. The calculation assumes that the rate of evolution of the periastron is much lower than the orbital frequency ($\dot\Phi\ll\dot\phi$). This allows us to expand
the true anomaly in terms of the Fourier series of the mean anomaly ($l$) with coefficients $S_n(e),\ C_n(e)$ and $A_n(e)$. The effects of periastron advance can be incorporated by replacing the constant $\Phi$ with its evolving counterpart. Note that in all the equations below, the eccentricity $(e)$, semi-major axis $(a)$, and mean anomaly $(l)$ are time-dependent quantities.

We write down the two polarisations as
\begin{widetext}
\begin{align} \label{eq:gw_strain_hc}
    h_\times &= -\frac{4m_1m_2}{a(1-e^2)}\frac{\cos i}{d_L} \sum_{n=1}^\infty \left[ \frac{S_n(e) - C_n(e)}{2} \sin\left(nl + 2\Phi\right) + \frac{S_n(e) + C_n(e)}{2} \sin\left(nl - 2\Phi\right) \right] \\
    h_+ &= -\frac{4m_1m_2}{a(1-e^2)}\frac{1}{d_L} \label{eq:gw_strain_hp}\sum_{n=1}^\infty \left[ A_n(e) \frac{\sin^2 i}{2} \cos nl + \frac{(1 + \cos^2 i)}{2} \left[ \frac{S_n(e) - C_n(e)}{2} \cos\left(nl + 2\Phi\right) + \frac{S_n(e) + C_n(e)}{2} \cos\left(nl - 2\Phi\right) \right] \right]
\end{align}
\end{widetext}

where the evolving periastron advance can be obtained by integrating the following equation
\begin{equation}
    \dot\Phi = \frac{3(GM)^{3/2}}{a^{5/2}(1-e^2)c^2} \label{eq:precession}
\end{equation}
with $M$ being the total mass of the binary.
Intermediate steps of the above derivation can also be found in Appendix \ref{appendix:theory}


Both the plus and cross polarisations have the same arguments of $nl\pm 2\Phi$. Thus, the \ac{gw} strain presents a number of tracks in the time frequency representation which encodes the time evolution of the binary system. We refer to them as higher harmonics. However, note that the plus polarisation has a term in the sum proportional to $\sin^2i$. For our analysis, we choose to ignore this term since it is largely suppressed unless the system is edge-on. Additionally, since it only appears in the plus polarisation depending on the source sky location, it will tend to be weaker than the other harmonics.

To determine the dominant harmonic in this Fourier expansion, we refer to \cite{Seto:2001pg}, which gives a Taylor series expansion for some of the $S_n$'s and $C_n$'s up to second order in eccentricity. Thus we have
\begin{align}
    S_1 &= C_1 + \mathcal O(e^3) = -\frac{3}{4}e + \mathcal O(e^3) \label{eq:S1} \\
    S_2 &= C_2 + \mathcal O(e^3) =  1-\frac{5}{2}e^2 + \mathcal O(e^3) \label{eq:S2} \\
    S_3 &= C_3 + \mathcal O(e^3) =  -\frac{9}{4}e + \mathcal O(e^3) \label{eq:S3}
\end{align}
It is clear that for low eccentricities, the $n=2$ term dominates over the other terms. 

We now wish to relate the frequency evolution of the individual terms in Eqs. \eqref{eq:gw_strain_hc} and \eqref{eq:gw_strain_hp} with the \ac{gw} frequency ($f$) averaged over one cycle. The Peters and Mathews \cite{Peters:1963ux} evolution equations provide the frequency and eccentricity evolution equations that can be numerically integrated under the Newtonian approximation to give the frequency as a function of time
\begin{equation}
    \left<\dv{f}{t}\right>=\frac{96\pi f^2}{5}(\pi \mathcal{M}f)^{5/3}\left(\frac{1+\frac{73}{24}e^2+\frac{37}{96}e^4}{(1-e^2)^{7/2}}\right) \label{eq:dfdt}
\end{equation}
\begin{equation}
    \left<\dv{e}{t}\right>=-\frac{304}{15}\frac{1}{\mathcal{M}}\frac{(\pi \mathcal{M}f)^{5/3}}{(1-e^2)^{5/2}}\left(1+\frac{121}{304}e^2\right) \label{eq:dedt}
\end{equation}
For the quadrupolar mode, the GW frequency is twice the average orbital frequency. 
Using Eqs. \eqref{eq:dfdt} and \eqref{eq:dedt}, one obtains the expression for $f(e)$ as \cite{Moore:2016qxz}

\begin{equation}
    \frac{f}{f_{\mathrm{ref}}} =
\left( \frac{e_{\mathrm{ref}}}{e(t)} \right)^{18/19}
\left( \frac{1 - e^2}{1 - e_{\mathrm{ref}}^2} \right)^{3/2}
\left( \frac{304 + 121 e_{\mathrm{ref}}^2}{304 + 121 e^2} \right)^{\frac{1205}{2299}}
\label{eq:fofe}
\end{equation}
with $f_\mathrm{ref}$ as the initial reference frequency and $e_\mathrm{ref}$ being the eccentricity at $f_\mathrm{ref}$. 

In the case of eccentric orbits, the orbital frequency has contributions from both; changes in the true anomaly and periastron precession. Thus, we can write down the following relationship between the \ac{gw} frequency, the average rate of change of the true anomaly and the rate of periastron advance as
\begin{equation}
    2\pi f = 2\left(\left<\dv{\phi}{t}\right> + \dot\Phi\right) \,.
\end{equation}
We can replace the average rate of the true anomaly by the rate of the mean anomaly if the time period of the binary does not vary appreciably over one orbital cycle. Thus we obtain
\begin{equation}
    \pi f = \dv{l}{t} + \dot\Phi \,.
    \label{eq:angle_rate_relations}
\end{equation}
From the phase evolution of individual eccentric harmonics, we write down the frequency evolution of $f_n$ as 

\begin{equation}
\label{eq:higherharm}
    f_n \equiv \frac{1}{2\pi}\dv{}{t}\left(nl\pm 2\Phi\right) = f + \frac{n\pm 2}{2\pi}\dv{l}{t} \,.
\end{equation}
Here, we explicitly use Eq. \eqref{eq:angle_rate_relations}. Thus, each harmonic is offset from the average \ac{gw} frequency by a quantity proportional to the rate of the mean anomaly. Further, using Eqs. \eqref{eq:S1}, \eqref{eq:S2} and \eqref{eq:S3} , one can show that the coefficients of the terms involving $(nl + 2\Phi)$ vanish at $\mathcal{O}(e^2)$. Thus, the fundamental track evolves as $f_2 \sim f$. Note that Eq. \eqref{eq:higherharm} is identical to Eq. (13) in \cite{Patterson2025} with $k\rightarrow n\pm 2$ and $f\rightarrow f_\phi$. We can also rewrite Eq. \eqref{eq:higherharm} with the time derivative of the mean anomaly explicitly substituted in, using Eq. \eqref{eq:angle_rate_relations}.  We use the fact that $f_2=f$ and neglect the subdominant frequency modes. Further, using Eq. \eqref{eq:precession} and Kepler's law third law, we get
\begin{align}
    f_n &= f_2 + \frac{n - 2}{2\pi}\left(\pi f_2 - \dot\Phi\right) \\
        &= f_2 \left[\frac{n}{2}- \frac{3}{2 (1-e^2)} \left(\frac{\pi G M f_2}{c^3}\right)^{2/3}\right]\,.
        \label{eq:ecc_harm}
\end{align}

The individual harmonics can then be obtained 
from Eqs. \eqref{eq:fofe} and \eqref{eq:ecc_harm}. With the theory of eccentric systems firmly in place, we now move to the systems we aim to target with this approach.

\subsection{Effective Chirp Mass Model: A phenomenological approach} \label{sec:ecmm_phenom}

In the previous subsection, we showed that the fundamental mode evolves as the $f_2$ mode. Thus, numerical solution of Eqs. \eqref{eq:dfdt} and \eqref{eq:dedt} provides $f(t)$ and $e(t)$, and thus the frequency evolution of the fundamental track. 
\citet{Bose2021} developed a phenomenological model for the $f_2$ time-frequency track as 
\begin{align}
	\frac{96\pi^{8/3}}{5}\left(\frac{G \mathcal{M}_e}{c^3}\right) ^{5/3} (t_c-t) - \frac{3}{8} f_2^{-8/3} = 0.
    \label{eq:fund_ev}
\end{align}
Using phenomenological fits, with the EccentricTD waveforms, the authors showed that in a restricted parameter space, $f_2(t)$ can be characterised by a phenomenological effective chirp mass ($\mathcal{M}_e$) (ECMM), which combines the chirp mass and the fiducial eccentricity $e_{10}$ (the eccentricity of the system at \ac{gw} frequency of 10 Hz) as given by
\begin{align}
	\label{eq:Me}
\mathcal{M}_{e} = \mathcal{M} (1 + \alpha (\mathcal{M}) e_{10}^2 + \beta (\mathcal{M}) e_{10}^4 + \gamma (\mathcal{M}) e_{10}^6)
\end{align}
where, $\alpha, \beta$ and $\gamma$ are polynomials in $\mathcal{M}$ (see \cite{Bose2021} for details). 

This work was extended in \citet{Hegde2024} with the incorporation of higher eccentric harmonics, allowing additional constraints to be placed in the $\mathcal{M}-e_{10}$ space. The authors proposed the frequency evolution of the first two eccentric harmonics as 
\begin{align}
f_3 &= k_3(\mathcal{M}, e_{10})f_2 \label{eq:f3}\\
f_4 &= k_4(\mathcal{M}, e_{10})f_2 \label{eq:f4}
\end{align}

The scale factors $k_3$ and $k_4$ were explicitly obtained in \cite{Hegde2024} by injecting eccentric systems (on $\mathcal{M}-e_{10}$ grid) into Advanced LIGO noise with an SNR of 1000 and searching for the maximum energy tracks in the corresponding time-frequency maps. In Sec \ref{sec:analytical} we show that using the scale factors satisfactorily approximates the analytical tracks given by Eq. \eqref{eq:ecc_harm}. 
\subsection{Detectability}

\begin{figure}
    \centering
    \includegraphics[width=\linewidth]{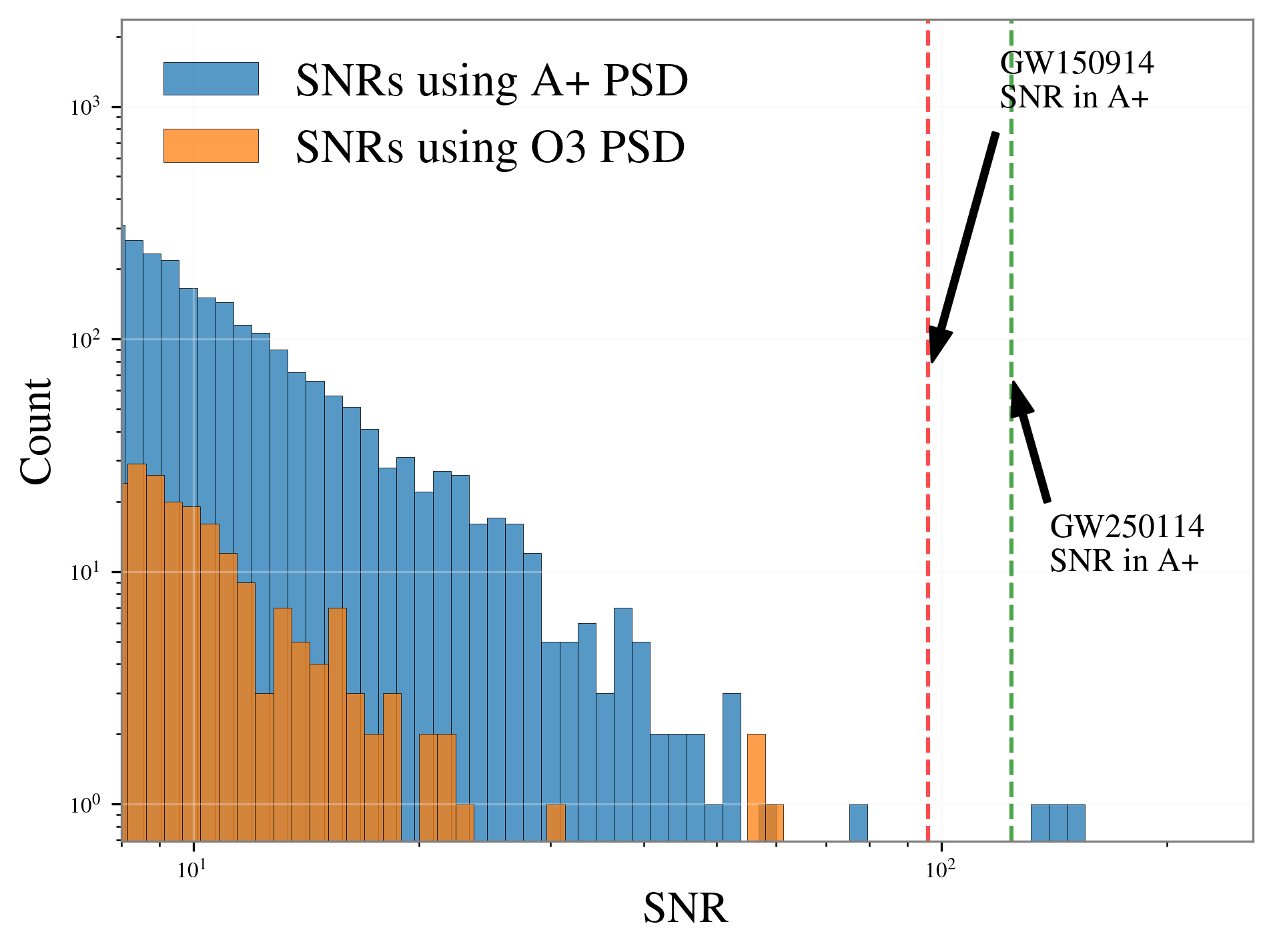}
    \caption{A histogram of SNRs for signals simulated with the GWTC-3 population parameters over an observation time of 1.5 years, assuming a merger rate density of $44\mathrm{Gpc^{-3}yr^{-1}}$}
    \label{fig:snr_hist}
\end{figure}

To assess the loudness of the eccentric BBH systems, we inject an astrophysical population of binary black holes consistent with the parameters of a power law plus peak distribution as obtained from the GWTC-3 population \cite{KAGRA:2021duu}. However, this does not take eccentricity into account, so we nominally set a uniform prior over the eccentricity, ranging from 0 to 0.6. We generate the waveforms using TEOBResums-DALI as it supports both precession as well as eccentricity and compute the optimal SNR of these signals using the A+ noise curve \cite{Albanesi:2025txj, Nagar:2024oyk, KAGRA:2013rdx}. We generate signals upto a redshift of 3 and assume a local merger rate density of $44\mathrm{Gpc^{-3}yr^{-1}}$, which is the upper merger rate density estimated using the GWTC-3 catalogue \cite{KAGRA:2021duu}. Fig \ref{fig:snr_hist} shows the histogram of the optimal single detector SNRs when computed with the O3 and A+ noise PSDs.  We note that GW150914 \cite{LIGOScientific:2016aoc} will be observed with a single detector SNR of 96 in A+ noise. The more recently detected loudest event, GW250114 \cite{LIGOScientific:2025rid} with an SNR of 80 will have an SNR of about 120 in an A+ sensitivity noise PSD. Thus, over a one and a half year period, we find that we would have three events with SNRs exceeding about 100. Thus, a choice of a single detector SNR of 100 looks reasonable in the A+ era.


Additionally, since we are going to be using the inspiral part of the signal to estimate the eccentricity (as described in the following section), we restrict ourselves to a detector frame chirp mass between $8M_\odot$ to $32M_\odot$ where the inspiral SNR will be significant. We also limit ourselves to equal mass systems in order to limit any effects of higher order spherical harmonics appearing from unequal mass BBH systems in the time frequency map. We will consider the case of unequal mass systems in a future work.

\section{Constraining Eccentricity} \label{sec:methodology}

\subsection{Review of Existing Phenomenological Approaches}
\begin{figure}
    \centering
    \includegraphics[scale=1]{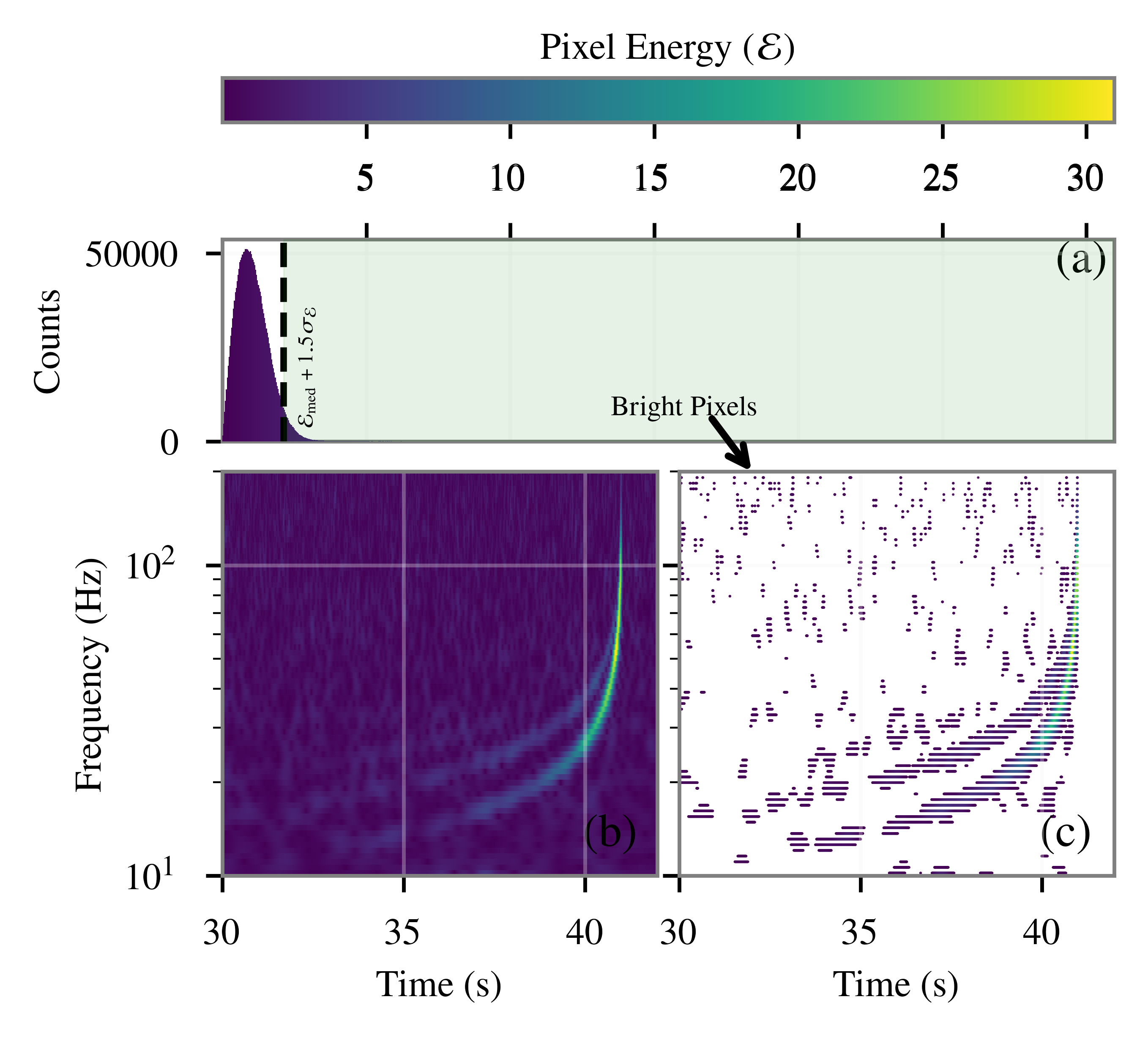}
    \caption{Schematic explaining the bright pixel extraction methodology. In subfigure (b), we have plotted the Q-transform of a \ac{gw} data, which is obtained by injecting a signal for an eccentric-\ac{bbh} system ($\mathcal{M}=15 M_{\odot}$, $e_{10}=0.2$) into Advanced LIGO noise with SNR=100. The subfigure (a) on the top plots the corresponding energy distribution of the pixels. The green patch in this figure denotes the bright pixels with an energy lower cutoff (denoted by the vertical black dashed line). The subfigure (c) plots the bright pixels separately to show the fundamental and first eccentric harmonic track.}  
    \label{fig:tfmap}
\end{figure}

\subsubsection{The Effective Chirp Mass Model}
\label{sec:ecmm}

In \cite{Bose2021, Hegde2024}, the authors had developed the \ac{ecmm} and proposed the idea of constraining the eccentricity by gridding over $\mathcal{M}-e_{10}$ space and recovering the higher harmonic tracks from the time-frequency representation using the Q-transform \cite{Brown1991}.

Given the Q-transform of the \ac{gw} data containing the injection, the authors attribute each grid point an importance dependent on the amount of pixel energy collected.
The energy recovery involves three steps illustrated by Fig. \ref{fig:tfmap}. In the first step, bright pixels with energies greater than median+1.5$\sigma$ of the pixel energies are identified where median is the median and $\sigma$ is the standard deviation of the pixel energy distribution.
The green patch in the inset of the leftmost plot in Fig. \ref{fig:tfmap} denotes the same. 
While collecting energy for a particular track, the authors collect not only the pixel energy closest to the track but also a few pixels above and below the track. These additional pixels are collected to account for detector noise and spectral leakage that occurs during the Q-transform. Among the collected pixels, the authors further throw away the pixels that are not bright. This is done for both the fundamental and the first harmonic tracks (which is computed using Eq. \eqref{eq:f3}). The mean pixel energy along these tracks is summed and normalised by the maximum value on the grid to obtain the normalised energy distribution in the $\mathcal{M}-e_{10}$ space. This distribution places bounds on the values of the chirp mass and eccentricity. 

\subsubsection{SNR Decomposition Based Constraints}

In a recent work ~\citet{Patterson2025} introduced a new framework for estimating eccentricity using higher eccentric harmonics. The authors aim to construct a collection of waveforms $h_k$, with each $h_k$ containing a harmonic with frequency $f_k$ given by Eq \eqref{eq:ecc_harm}. To identify these $h_k$ they perform a Singular Value Decomposition (SVD) on a collection of eccentric waveforms $x_j$ with all parameters fixed except the mean anomaly. They are able to show that the leading four singular vectors can be identified with the eccentric harmonics $h_k$ with $k=0,-1,1,2$ which corresponds to $n = 2,1,3,4$ with frequencies $f_n$ in our notation. With this identification, they infer that the $h_k$'s can be generated as
\begin{align}
    h_{k} = \frac{1}{N} \sum_{j=0}^{N-1} e^{\left(2\pi ijk/N\right)} x_{j}
    \label{eq:ben_ecc_harm}
\end{align}
where $N$ is the number of  eccentric waveforms ($x_j$) used with $N=6$. This allows for a computationally efficient method to obtain the harmonics $h_k$. An additional Gram-Schmidt orthogonalisation is used to make the $h_k$'s orthogonal.


Subsequently, the authors obtain the curve along which $h_0$ is degenerate in the chirp mass - eccentricity plane.
They find that for any given eccentric waveform, its overlap with the $k=0$ harmonic achieves high values along the line of degeneracy. However, this line of degeneracy is not respected by the subdominant harmonics 
allowing the authors to estimate the eccentricity using the SNR ratios of the different harmonics with the $h_0$.

Given an eccentric signal, the authors obtain the eccentricity as follows. A quasicircular \ac{pe} is performed, giving an estimate of the chirp mass that differs from the true chirp mass due to the system's eccentricity. The degeneracy curve in the chirp mass and eccentricity space passing through this chirp mass estimate is calculated, and the $x_j$'s are constructed at a point on this degeneracy curve. These are used to obtain the $h_k$'s that matched with the data to obtain the respective SNRs $\rho_k$ for $k=0,1,-1$. The ratios of these SNRs are used to estimate the eccentricity.

\subsection{Methodology}

Here, we extend the framework introduced in \cite{Bose2021, Hegde2024} to constrain the eccentricity of the BBH source. We propose several improvements to the methodology, which we will discuss in this section.

\subsubsection{Energy-informed approach in energy collection }\label{sec:EIEC}

To recall, for faithful extraction of the tracks from the Q-transform of the data, after extracting the bright pixels, we need to select the relevant pixels along the frequency track for energy collection. As explained before in section \ref{sec:ecmm}, \cite{Bose2021, Hegde2024}, the authors used a fixed pixel width approach in which they collected pixels that are closest to the track and collected additional $2$ ($1$ for the higher eccentric harmonic tracks) pixels from each side of the frequency of the pixel to account for the spectral leakage.

In this work, we use additional information of the energy profile along the frequency bins at a given temporal location to retain the relevant pixels. We argue that, for each time slice, the track corresponding to the injected signal would be closest to the pixel with maximum energy. Therefore, for a given time slice, the energy of the pixel closest to the chosen track gives the maximum energy pixel. Thus, from the selected pixels following a fixed-width prescription, we discard all pixels with energy higher than the energy of the pixel closest to the track. Thus, we do not allow unnecessary pixels that are more energetic; therefore, we reduce the redundancy between nearby time-frequency tracks in the sample space.
\begin{figure}
    \centering
    \includegraphics[scale=1]{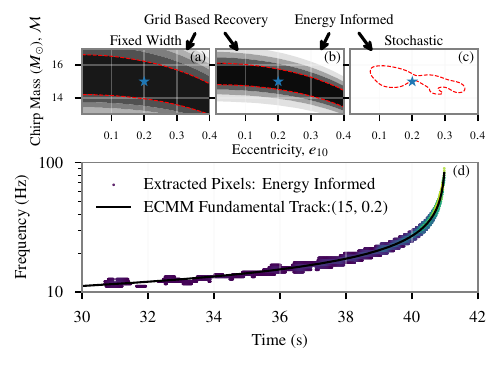}
    \caption{Comparison of the performance of the new pixel extraction and stochastic sampling recovery methodology. Subfigure (d) plots the pixels extracted from bright pixels obtained in Fig. \ref{fig:tfmap} using the energy-informed method for the fundamental mode. It also plots the fundamental track for the injected ($15 M_{\odot}$, $0.2$) system (denoted by black solid line) using ECMM. The three subfigures on top plot the recovery contours for the injected system using (a) Fixed-Width pixel extraction with grid-based recovery, (b) Energy Informed pixel extraction with grid-based recovery, and (c) Energy Informed pixel extraction with stochastic sampling-based recovery. The blue star denotes the injected ($15 M_{\odot}$, $0.2$) system.}
    \label{fig:colenerg}
\end{figure}

In Fig. \ref{fig:colenerg}, we plot the pixels collected by the fundamental track of the injected system ($\mathcal{M}=15$, $e_{10}=0.2$) using the energy-informed methodology. 
In the insets of Fig. \ref{fig:colenerg} (a) and (b), we plot the normalised energy contours for both the energy collection methods, with a fixed number of pixels and energy-informed pixel selection. We find that the energy-informed pixel collection approach reduces the area within the 0.95 contour level, thereby reducing redundancy between nearby tracks. 

\subsubsection{Stochastic Sampling}
Along with the improvement in the pixel extraction method we also introduce a likelihood based sampling approach inspired by Bayesian parameter estimation.
The parameter constraining approach used in \cite{Bose2021, Hegde2024} involves collecting energy for BBH systems with parameters chosen from a rectangular grid in a $\mathcal{M}-e_{10}$ space and normalising this energy to define the distribution $p(\mathcal{M}, e_{10} |d)$ where $d$ corresponds to the time-frequency map of the \ac{gw} data. To improve the accuracy of $p(\mathcal{M}, e_{10} |d)$, we can either increase the density of the $\mathcal{M}-e_{10}$ grid or rely on a more efficient sampling in $\mathcal{M}-e_{10}$ space. As usual, the best strategy is to sample more points in the neighbourhood of the maximum energy value. Therefore, we use the Bayesian framework to write 
\begin{align}
    p(\mathcal{M}, e_{10} |d) \propto p(d|\mathcal{M}, e_{10}) p(\mathcal{M}, e_{10}).
\end{align}
Here, $p(\mathcal{M}, e_{10})$ is the prior and $p(d|\mathcal{M}, e_{10})$ is the likelihood, $\mathcal{L}$, that the data contains the proposed BBH system with the given parameters. We argue that the collected energy, E, is proportional to the likelihood, i.e., $\mathcal{L} = \kappa E$. Here $\kappa$ is the normalisation constant. Since our interest is in $p(\mathcal{M}, e_{10} |d)$, which is proportional to the $\mathcal{L}$, henceforth, we will drop the proportionality constant.

\subsubsection{Exploring likelihood choices} \label{sec:likelihood_choices}

In the previous subsection, we argued that the energy collected along a given track can be used to construct a likelihood. In \cite{Hegde2024}, the authors demonstrated that including the additional energy from the first eccentric harmonic can significantly improve constraints on the eccentricity. Specifically, they chose to maximize the combined energy collected along the fundamental track, $E_2$, and the first harmonic, $E_3$ to enhance the eccentricity estimation.

Here, we explore two possibilities, namely taking the product of the energies or a square of their sums \footnote{Here, we choose a square of the sums to make the unnormalised likelihood have the same dimensions as the product.} denoted by $\mathcal{L}^\pi$ and $\mathcal{L}^\sigma$ respectively. We note that at low eccentricities, the first harmonic contains negligible energy, making $\mathcal{L}^\pi\rightarrow 0$. This is not the case for $\mathcal{L}^\mathrm{\sigma}$ since at low eccentricities it is dominated by $E_2$. To remedy this, we construct a piecewise likelihood given below such that at low eccentricities, we exclude $E_3$ in the likelihood. This choice is governed by the threshold energy ratio $\alpha$.
\begin{equation}
  \mathcal{L}^\pi = 
    \begin{cases}
        E_2 E_3 & E_3^*/E_2^*\geq \alpha \\
        \alpha E_2^2 & E_3^*/E_2^*< \alpha \\
    \end{cases}
\end{equation}

To keep the comparison consistent between $\mathcal{L}^\pi$ and $\mathcal{L}^\sigma$, we opt to apply a similar threshold to $\mathcal{L}^\sigma$ as well
\begin{equation}
  \mathcal{L}^\sigma = 
    \begin{cases}
        (E_2+E_3)^2 & E_3^*/E_2^*\geq \alpha \\
        [(1+\alpha) E_2]^2 & E_3^*/E_2^*< \alpha \\
    \end{cases}
\end{equation}

To fix $\alpha$, we perform a toy exercise with a very high SNR (of 1000) such that the noise realisation does not lead to a bias. We precompute the recovered energies over a grid of chirp mass and eccentricity values in the validity region of the ECMM and cache the ratio of the energy in the first harmonic to that of the fundamental track. During the likelihood sampling stage for any given sample, we estimate its energy ratio by performing a radial basis interpolation over the three nearest points. We denote this value by $E_3^*/E_2^*$, and $\alpha$ is a threshold that we set to 0.15.

At low eccentricities, we expect $\mathcal{L^\pi}$ to outperform $\mathcal{L^\sigma}$ based on the following argument. When the eccentricity is small and the first harmonic is effectively absent, $\mathcal{L}^\pi$ is unlikely to pickup points where $E_3^*/E_2^*\geq \alpha$ and $\mathcal{L}^\pi\sim 0$. $\mathcal{L}^\sigma$, on the other hand, will be dominated by the energy collected by the fundamental track, leading to $\mathcal{L}^\sigma \sim E_2^2,$ leading to poorer constraints. We verified this by performing an injection study, which showed that the product likelihood was able to recover more signals compared to $\mathcal{L^\sigma}$. Thus, in the following sections, we will use the product likelihood for our analyses.



Further, in addition to the energy content, we use additional information on the energy ratio between the first harmonic and the fundamental track. We append this into the likelihood by penalising samples whose collected energy ratio deviates from the cached energy ratio $E_3^*/E_2^*$. Thus, we further modify the likelihood as 

\begin{equation}
    \mathcal{L^\pi_\mathrm{pen}} = \mathcal{L^\pi}\ e^{-|E_3/E_2-E_3^*/E_2^*|} \label{eq:penalty_likelihood}
\end{equation}
The exponential ensures that the likelihood is penalised should $E_3/E_2$ deviate drastically from $E_3^*/E_2^*$. Note that one could impose stricter penalties in the form of a function that drops off more sharply than the exponential; however, we find that lower SNRs reduce the reliability of the energy ratio collected, and thus we opt not do so.


\section{Simulations and Results} \label{sec:results}

In this section, we summarise the results from a simulation study to assess the eccentricity constraints using the likelihood expression given in Eq. \eqref{eq:penalty_likelihood}.
\subsection{Non-spinning systems}

We construct an injection set of 500 signals in $\mathrm{A+}$ noise with an SNR of 100 with the inspiral-only EccentricTD
waveform for equal mass systems. To construct the priors, we use the fact that the chirp mass is weakly biased when recovering eccentric systems with quasicircular waveforms \cite{Divyajyoti:2023rht}, we restrict the chirp mass prior to $\pm 3M_\odot$ around the injected value. This range ensures that we are able to incorporate any uncertainties in point estimates of chirp mass from alternate methods, such as from quasicircular templated searches. Our eccentricity prior spans the range from 0 to 0.4. We also restrict the prior to be within the validity region of the ECMM. To compare the results and assess the performance, we construct the posterior contours at fixed percentile levels\footnote{Since the prior is uniform, the likelihood and posterior are identical within the validity region upto a normalisation factor}.
Here, if we uniformly sample from the bounding box of the points obtained from the stochastic sampling, then a p-th percentile contour level contains
p fraction of points within that contour. 

\begin{figure}
    \centering
    \includegraphics[width=\linewidth]{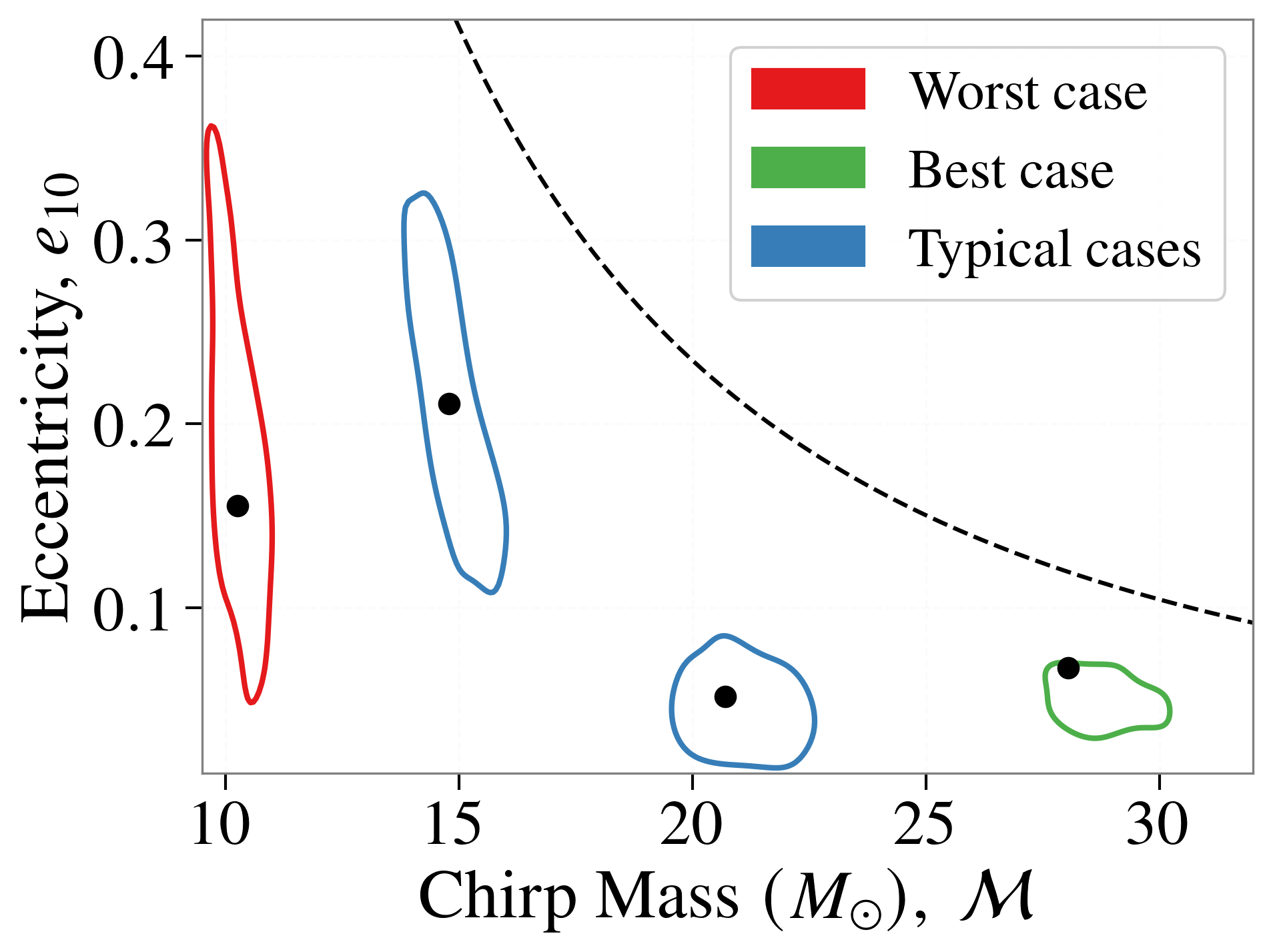}
    \caption{90th percentile contours for a few injections as recovered with the product likelihood and a restricted chirp mass prior along with the injected parameters. The points within the contours mark the injected values}
    \label{fig:prod_contours}
\end{figure}

\begin{figure}
    \centering
    \includegraphics[width=\linewidth]{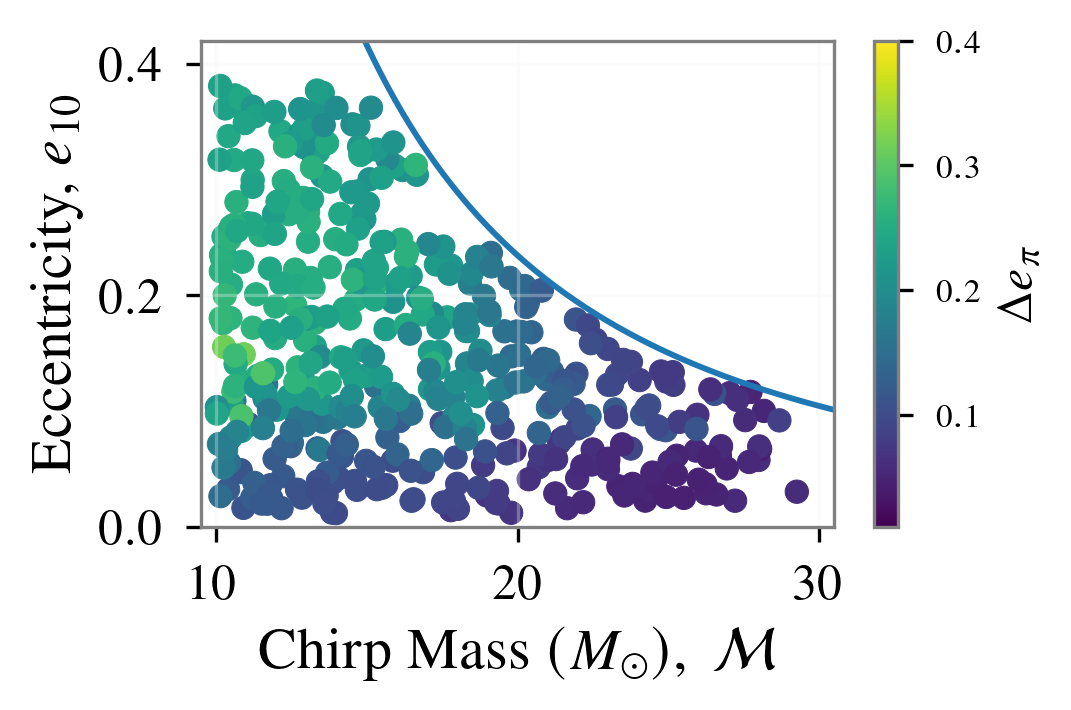}
    \caption{This figure plots 500 systems injected in A+ noise at an SNR of 100 and shows the eccentricity constraint as recovered by the product likelihoods for a known chirp mass. We discard systems where the injected value lies outside the 90th percentile contour.}
    \label{fig:delta_ecc_prod_thresh_scatter}
\end{figure}

We find that the 90th percentile contour works well for our purposes, as it gives good constraints while enclosing the injected chirp mass and eccentricity values. The extent of the contours along the eccentricity direction ($\Delta e$) will be our metric for the performance. This is a proxy for the constraint on the eccentricity. 

In Fig \ref{fig:delta_ecc_prod_thresh_scatter} we plot $\Delta e$ as a function of the injected chirp mass and eccentricity. We have discarded systems for which the injected parameters lie outside the 90th percentile contour, as they lie at the edge of the validity region leading to pathological contours. We obtain a median $\Delta e$ of 0.17, indicating that on average we are able to provide reasonable, coarse constraints on the eccentricity if the chirp mass is known. The best constraint for an injected system within the 90 percentile contour is obtained for a binary with a chirp mass and eccentricity of $28.04M_\odot$ and 0.07, respectively. For this system, we find a $\Delta e$ of 0.04. We show the 90th percentile contour for this system in Fig \ref{fig:prod_contours}. The tight constraint is partly due to the validity region effectively placing an upper bound on the eccentricity. Fig \ref{fig:prod_contours} also shows our worst-case scenario, which lies at the extreme lower end of the chirp mass with a low eccentricity. This is due to the fact that the signal has a weak first harmonic track, which forces us to rely on the fundamental track itself, which is unable to constrain the eccentricity due to the chirp mass eccentricity degeneracy (see Appendix \ref{appendix:ecmm}). However, despite this, it is clear that our method can provide reasonable constraints on the eccentricity.

We also test the recovery of the systems with differing SNRs. It would be reasonable to expect that with higher SNRs, we obtain tighter constraints. This is, however, not the case as we are limited by the spectral leakage of the Q transform kernel. Increasing the SNR does not produce sharper tracks, limiting our ability to tightly constrain the eccentricity. This constancy of $\Delta e_\pi$ also shows up at lower SNRs. However, below an SNR of 15, we are unable to reliably constrain the eccentricity as the tracks are diffuse, and bright pixels cannot be reliably identified.



\begin{widetext}
\begin{figure*}[ht!]
   \centering
        \begin{subfigure}[t]{0.45\textwidth}
           \includegraphics[width=\textwidth]{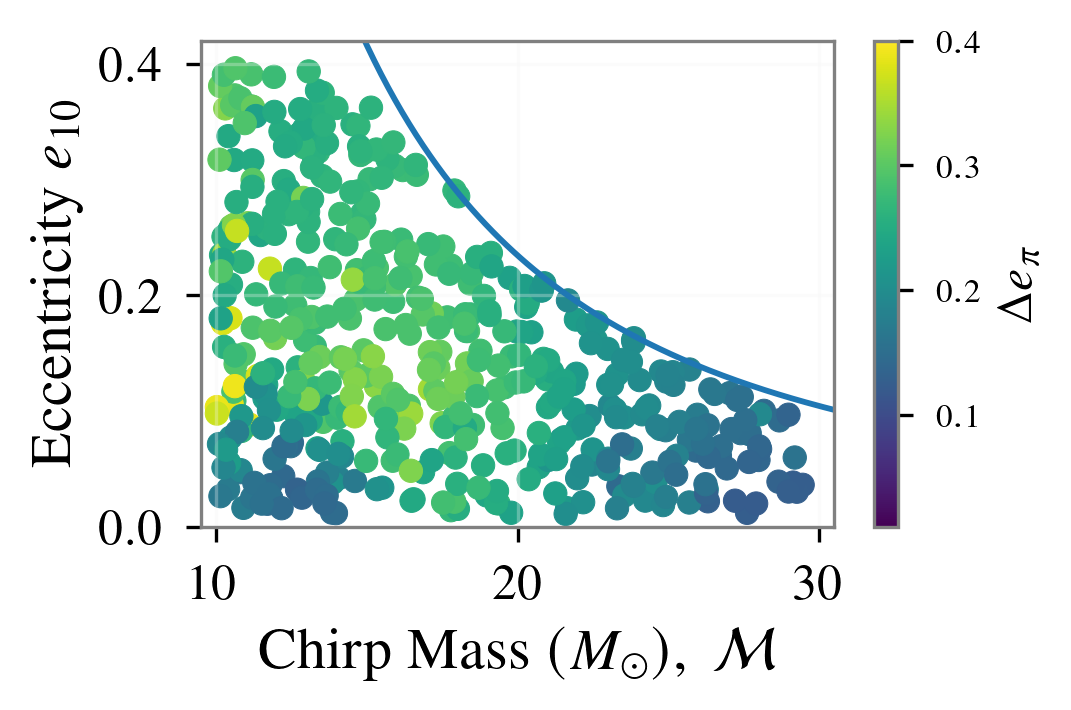}
           \caption{Uncertainty in the eccentricity for aligned spin systems with $s_{1z}=s_{2z}=0.1$}
           \label{subfig:delta_ecc_spin01}
        \end{subfigure}%
\hspace{1cm}
        \begin{subfigure}[t]{0.45\textwidth}
           \includegraphics[width=\textwidth]{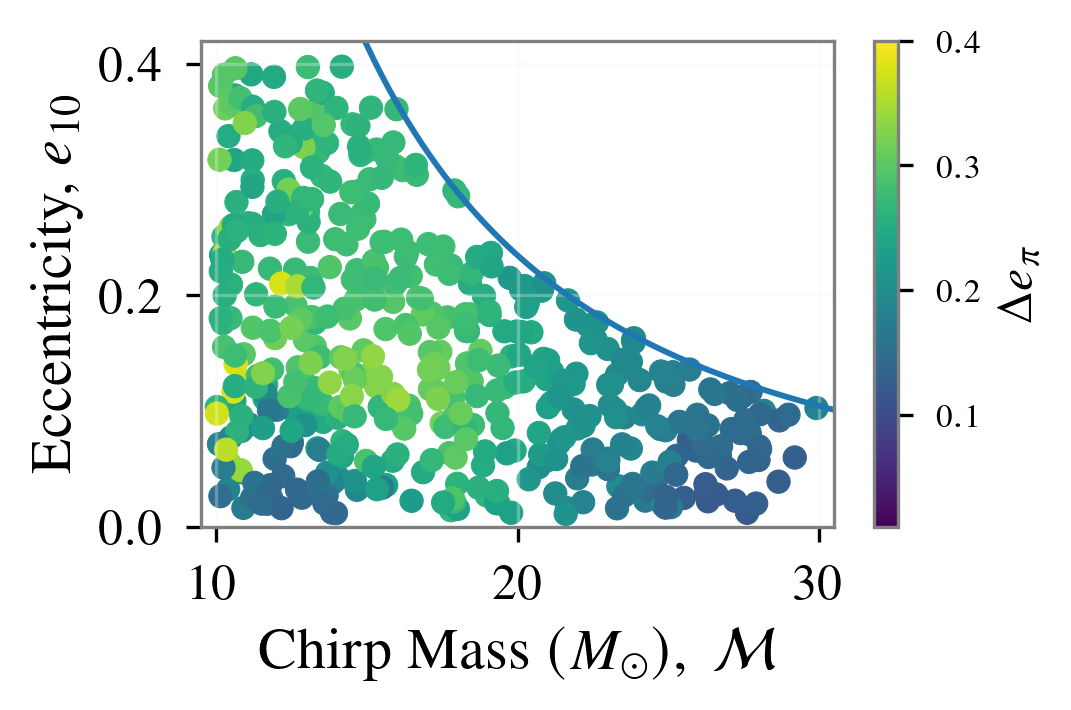}
           \caption{Uncertainty in the estimated eccentricity for aligned spin systems with $s_{1z}=s_{2z}=-0.1$}
           \label{subfig:delta_ecc_spinneg01}
        \end{subfigure}
\hspace{1cm}
\caption{Uncertainty in recovery for aligned spin systems}
\label{fig:delta_ecc_spins}
\end{figure*}
\end{widetext}

\subsection{Aligned spin}
In this section, we attempt to test the proposed approach for spinning systems, restricting ourselves to BH spins aligned to the orbital angular momentum with both BHs having equal spins. We generate our injections with the TEOBResumS-Dali model, which also supports eccentric, precessing systems. 
We recompute the energy ratio bookkeeping, which keeps track of the energy ratio information for this waveform model. These energy ratio values are then used in the piecewise likelihood detailed in \ref{sec:likelihood_choices}. 

We consider 500 equal mass systems with eccentricities within the validity region and fixed spin values of $s_{1z}=s_{2z}=\pm0.1$ and inject them with an SNR of 100 in $\mathrm{A}+$ noise. As before, we show the $\Delta e$ values for both the aligned and anti-aligned spins in Fig. \ref{fig:delta_ecc_spins}. The $\Delta e$ trend is preserved, with the best constraints being obtained at the lowest eccentricities. For positive aligned spins, the contour contains 96\% of systems, while for anti-aligned spins, about 98\% are contained in the 90th percentile contour, indicating that our model is robust towards aligned spins. 
Spins above 0.1 give poor results as the signal length increases, which is partly degenerate with lower chirp mass values. Here, we do not impose the restricted chirp mass prior since our model is biased towards low chirp masses. This is due to the fact that we are recovering injections made with TEOBResumS as opposed to EccentricTD, which was what our method was tuned with. We find that recovering TEOBResumS injections tends to show a bias towards lower masses in the recovered energy for the same injected parameters and thus the contours are not centred around the injected chirp mass value (see appendix \ref{appendix:systematics}). As a result, reporting the contour extents for the injected chirp mass would be very small, which would artificially indicate tight constraints. For both positive and negative aligned spins, we find a median $\Delta e$ of 0.23, indicating that our method is not significantly affected by spins within our tested range.


\section{Comparison of the scale-factor based model with the analytical Eccentric Harmonics} \label{sec:analytical}

In this section, we compare the eccentric harmonic tracks obtained using the scale factor Eqs. \eqref{eq:f3}, \eqref{eq:f4} with the ones obtained analytically Eq. \eqref{eq:ecc_harm}.
We recall that in Patterson et al. \cite{Patterson2025}, the authors compute the fundamental track from the time frequency map, obtaining the brightest track and higher harmonics by solving Eqs. \eqref{eq:fofe} and \eqref{eq:ecc_harm}. In Hegde et al \cite{Hegde2024}, authors developed the scale factor model. We qualitatively assess the frequency evolution from these two approaches for the eccentric harmonics.

\begin{figure}
    \centering
    \includegraphics[scale=1]{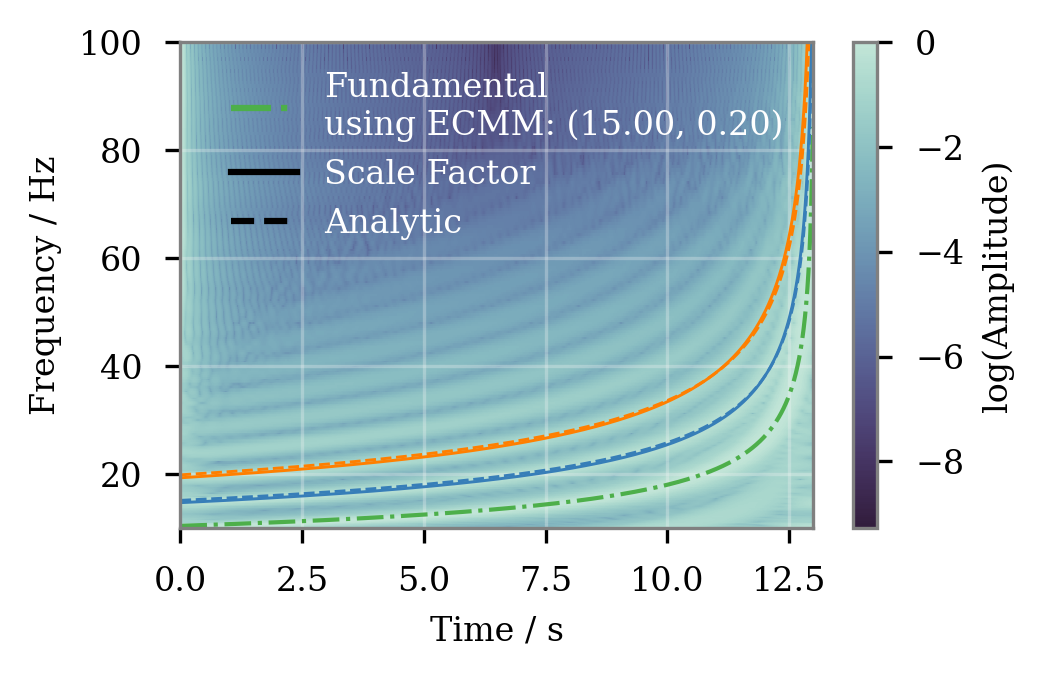}
    \caption{Time frequency map of the plus polarisation of gravitational waveform coming from eccentricTD model for \acp{bbh} with $(q, \mathcal{M}/M_{\odot}, e_{10}):(1, 15, 0.2)$. Also plotted are frequency evolution track corresponding to fundamental (2, 2) mode (dot-dashed green line) obtained from ECMM, and the first two eccentric harmonic tracks (blue and orange respectively) obtained by multiplying fundamental track with the scale factors (solid line) or by using analytical expression given in Eq. \eqref{eq:higherharm} (dashed line)}
    \label{fig:benvsscale}
\end{figure}

In Fig.~\ref{fig:benvsscale}, we plot the Q-transform of the plus polarisation of GW waveform using the eccentricTD model that corresponds to $(q, \mathcal{M}/M_{\odot}, e_{10}):(1, 15, 0.2)$ system. We then use the ECMM (Eq. \eqref{eq:Me} and Eq. \eqref{eq:fund_ev}) to obtain the fundamental track ($f_2 (t)$, denoted by green dot-dashed line in Fig. \ref{fig:benvsscale}).

Next, we use scale-factor-based and analytical expression-based methods to obtain the evolution track for the first two eccentric harmonics. The first method, developed in \citet{Hegde2024}, as described in Sec \ref{sec:ecmm_phenom} obtains the $nth$ ($n\in {3,4}$) harmonic by multiplying the fundamental track (i.e. $n=2$) by a scale factor, $k_n (\mathcal{M}, e_{10})$, denoted by solid blue (first harmonic) and orange (second harmonic) curves in Fig.~\ref{fig:benvsscale}. The other method~\cite{Moreno-Garrido1995,Patterson2025}, uses an analytic expression for the $k^\mathrm{th}$ ($k \in{1,2}$ i.e. $k=n-2$) eccentric harmonic, given by Eq. \eqref{eq:ecc_harm}. Note that in Fig \ref{fig:benvsscale}, we use the fundamental track as given by the ECMM model and not as obtained in Fig. 2 of Patterson et al.
The root of Eq. \eqref{eq:fofe} at each time slice gives the eccentricity evolution which starts from reference eccentricity, $e_{\text{ref}}$, at the reference frequency $f_{\text{ref}}$ (10 Hz in our case). The scale factor-based tracks are solid lines, and analytic expression-based tracks are dashed lines in Fig.~\ref{fig:benvsscale}. For this particular system, we find that the scale factor-based tracks follow closely with the analytic tracks.


\begin{figure*}
    \centering
    \includegraphics[scale=1]{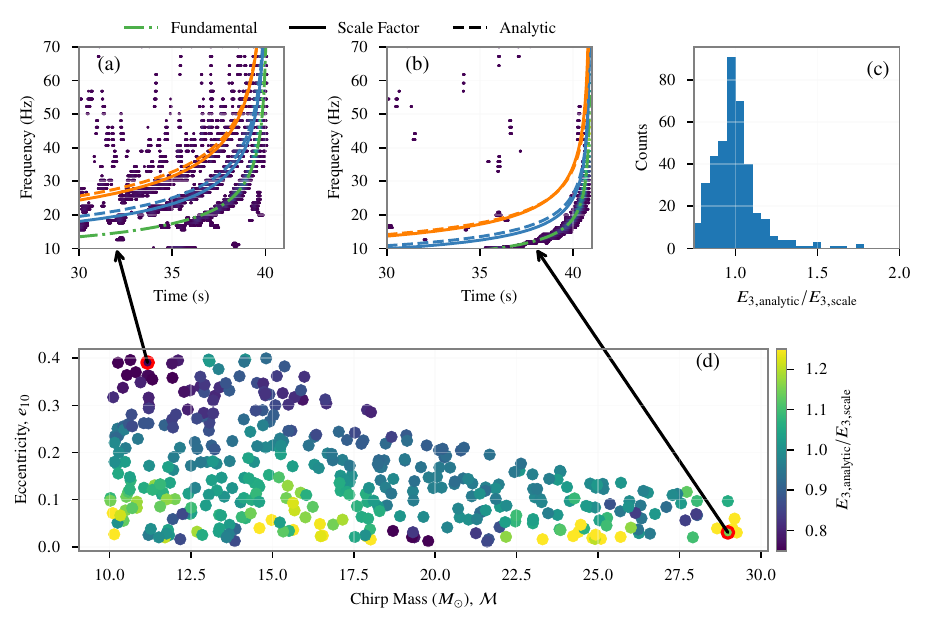}
    \caption{Comparing the energy collected (for the first eccentric track) using the scale factor with the energy collected using the analytical expression. Bottom panel (d) shows the contour plot for the ratio of the energy collected by frequency evolution track corresponding to the first eccentric harmonic obtained through the scale factor method and analytical method for a range of eBBH systems injected in advanced LIGO noise with SNR = 100. We also plot the corresponding bright pixels extracted for two representative eBBH systems (panels (a) and (b)). Subfigure (c) plots the distribution of the above energy ratio over the validity region.}
    \label{fig:E3_benvsscale}
\end{figure*}

To further test the efficiency of the energy extraction using the analytic expression, we inject a gravitational waveform (using EccentricTD model) for 400 eBBH systems, with $q=1$ and varying $\mathcal{M}, e$, in $\mathrm{A}+$ noise. 
We then compute the energy extracted (using energy informed method) by the injected systems using the first harmonic track ($E_3$) computed through the above two methods. In Fig.~\ref{fig:E3_benvsscale}, we plot the ratio of the energy collected from the first harmonic, $E_{3, \text{analytic}}/E_{3, \text{scale}}$, for these 400 systems. The bottom panel, Fig.~\ref{fig:E3_benvsscale}(d), shows the scatter plot for the chosen systems where the colour encodes the energy ratio. We also plot the Q-transforms (after extracting the bright pixels) and corresponding eccentric tracks for two systems in top left (Fig.~\ref{fig:E3_benvsscale}(a)) and top right (Fig.~\ref{fig:E3_benvsscale}(b)) panels. Fig.~\ref{fig:E3_benvsscale}(a) shows the tracks for eBBH system for which the scale factor based method collects more energy than the analytical one, while Fig.~\ref{fig:E3_benvsscale}(b) corresponds to the system where the situation is reversed. Finally, Fig.~\ref{fig:E3_benvsscale}(c) plots the histogram of the energy ratio. We find that the energy ratios are largely close to one (i.e. $0.98^{+0.26}_{-0.18}$), which means that the tracks obtained from the two methods are consistent. However, for low $\mathcal{M}$, high $e_{10}$ regions, the ECMM fundamental track deviates from the EccentricTD track (as shown in Fig. \ref{fig:E3_benvsscale}(a)). This leads to deviation in the analytic first harmonic track. However, the scale factor based track by design in the phenomenological model is expected to collect more energy than the analytical track. For the low $e_{10}$, high $\mathcal{M}$ region, the energy in first harmonics is low, and the variation of scale factor is high. Therefore, the analytic first harmonic track collects much more energy than the scale factor based track that leads to high ratios (as shown in Fig. \ref{fig:E3_benvsscale}(b)).

We repeat this analysis with the TEOBResumS-DALI waveform model  for non-spinning eccentric BBHs and find that this model is consistent with the ECMM for high eccentricity systems 
and therefore the ratio of the energy collected using the first two harmonic tracks approaches 1. However, the trend for low $e$, high $\mathcal{M}$ region persists. In summary, the energy collected from the first harmonic track using the scale factor based method is close enough to the ones obtained using the analytic expression. Therefore, we can replace the scale factor based higher harmonics track calculation with analytic ones. 
As the analytic expression is a function of mass ratio, we may be able to extend this formulation to asymmetric systems, which we plan to explore in more detail in the future.

\section{Conclusions}
 The current detections with the advanced interferometers have not so far, shown any confident signature of eccentricity. Given that waveforms capturing the complete physics of an eccentric \ac{bbh} system are still under development current matched filter searches do not incorporate the eccentric waveforms. Eccentric searches employed by the \ac{lvk} collaboration solely rely on the morphological, model-agnostic searches for eccentric signatures in the GW from the incoming \ac{cbc} sources. Alternative approaches have been explored such as \cite{Bose2021, Patterson2025} to develop phenomenological models to constrain the eccentricity of the underlying signal.

In \cite{Bose2021, Hegde2024}, the phenomenological effective chirp mass model and the corresponding scale factor was developed for eccentric harmonics using the time-frequency representation. These harmonics appear as separate tracks in a Q-transform with their frequency evolution dependent on the eccentricity and chirp mass at leading order. The authors show that their phenomenological model can provide reliable eccentricity constraints. Such methods provide an alternative approach to eccentricity estimation which we build upon in this work. 

We show improvements obtained by refining the pixel extraction algorithm and using a likelihood-based sampling. The improved pixel extraction method avoids double-counting the energy as the tracks get closer at late inspiral. Additionally, the likelihood formulation enables us to use stochastic samplers that are standard in Bayesian parameter estimation. We compare two likelihood formulations given the energies collected by the different tracks. We find that in the case of moderate eccentricity where systems display both, the fundamental and first harmonics, the product of the energies collected performs better than the squared sum of the energies. Our likelihood construction naturally extends to include an arbitrary number of higher harmonics. 

In addition, we also incorporate the crucial information of the energy ratio between the eccentric harmonics to further constrain the eccentricity. We demonstrate the ability of the thresholded likelihood with an energy ratio penalty to provide tight constraints on the eccentricity, with the best results obtained for lower values of the eccentricity. 

The likelihood-based method developed in this work is computationally efficient and can provide eccentricity estimates in about 5 minutes when run in parallel on a 50 core CPU. In case of the detection of a non-negligible eccentricity, it allows us to quickly identify interesting \ac{gw} candidates that formed via dynamical channels, allowing us to follow up events with potential EM counterparts, as can be expected from certain dynamical channels such as \ac{agn} assisted mergers \cite{Tagawa:2023uqa}. Our approach is robust to SNR variation, provided the eccentricity tracks are distinct in the time-frequency representation. Diffuse tracks arising from moderate inspiral SNRs (below 15) prevent accurate bright-pixel extraction, rendering eccentricity recovery impossible. 


However, our results are contingent on the injections being recovered with the same waveform model used for the phenomenological model. Moreover, as the phenomenological model was tuned to the non-spinning EccentricTD waveform, our recovery can be biased for spinning binaries that exhibit spin magnitudes exceeding 0.1. Waveform systematics can further introduce biases in certain regions of the parameter space. These biases can be reduced by tuning the phenomenological model to more accurate waveform models. Thus, this approach can be readily adapted, improving the reliability of our results. This can help in obtaining coarse level constraints on the eccentricity in a computationally efficient way which can provide an independent consistency check with a full Bayesian parameter estimation study. Finally, we show that we can move away from an empirical estimate of the frequency evolution of the higher harmonics to a more robust analytic method. This allows us to generalise our approach to unequal-mass systems, which we plan to tackle in an upcoming work. Such phenomenological approaches are crucial in the advent of next generation detectors, where the high SNRs and longer duration signals make it computationally expensive to run a brute-force parameter estimation pipeline.

\section*{Acknowledgements}
The authors thank Shubhanshu Tiwari and Shubhagata Bhaumik for their comments on the paper. JF acknowledges support from the Prime Minister’s Research Fund. PT acknowledges support from IIT-Bombay. AP acknowledges the support from SPARC MoE grant SPARC/2019-2020/P2926/SL, Government of India. This material is based upon work supported by NSF's LIGO Laboratory which is a major facility fully funded by the National Science Foundation.

\appendix
\section{Deriving the gravitational wave strain for an eccentric system} \label{appendix:theory}
Directly applying the quadrupole formula for \ac{gw} emission leads to
\begin{align}
h_\times &= -\frac{2 m_1 m_2}{a (1 - e^2)} \frac{2}{d_L} \cos i \left[\left(\frac{5e}{4} \sin \phi + \sin 2\phi + \frac{e}{4} \sin 3\phi\right) \right. \nonumber \\
& \quad \times \cos 2\Phi - \left.\left( \frac{5e}{4} \cos \phi + \cos 2\phi + \frac{e}{4} \cos 3\phi + \frac{e^2}{2} \right) \sin 2\Phi \right], \\
h_+ &= \frac{2 m_1 m_2}{a (1 - e^2)} \frac{1}{d_L} \sin^2 i \left(\frac{e^2}{2} + \frac{e}{2}\cos \phi \right) - \frac{1}{2} \nonumber \\
& \quad \times \left(\frac{1 + \cos^2 i}{\cos i}\right) h_\times \left(\Phi + \frac{\pi}{4}\right),
\end{align}

Note the polarisations above are trigonometric functions of the true anomaly $\phi$. These functions can now be decomposed in a Fourier series expansion of the mean anomaly $l$. We do not provide the expansion terms here but merely present the resulting series expression. 

\begin{align}
    h_\times &= -\frac{4m_1m_2}{a(1-e^2)}\frac{1}{d_L}\cos i \sum_{n=1}^\infty ( S_n(e) \sin nl \cos 2\Phi - \nonumber \\ 
           & \qquad \qquad \qquad \qquad \qquad \qquad C_n(e)\cos nl \sin 2\Phi) \\
    h_+ &= -\frac{2m_1m_2}{a(1-e^2)}\frac{1}{d_L}\sin^2 i \sum_{n=1}^\infty [(1-e^2)J_n(ne)\cos nl] - \nonumber \\
    & \qquad \frac{1}{2}\frac{1+\cos^2i}{\cos i}h_\times\left(\Phi+\frac{\pi}{4}\right) \label{eq:plus cross strain}
\end{align}

Here $J_n(ne)$ is the Bessel function of order $n$ and $S_n(e)$ and $C_n(e)$ are constructed out of a linear combination of Bessel functions and their first derivatives.  

Eq. \eqref{eq:plus cross strain} can be massaged into 
\begin{align}
    h_\times = &-\frac{4m_1m_2}{a(1-e^2)}\frac{\cos i}{d_L} \sum_{n=1}^\infty \left\{ \frac{S_n(e)-C_n(e)}{2} \nonumber \right. \\
    &\left. \times \sin[(nl+2\Phi)]+ \frac{S_n(e)+C_n(e)}{2}\sin[(nl-2\Phi)]\right\}  \label{eq:strain cross}
\end{align}
\begin{align}
    h_+ = &-\frac{2m_1m_2}{a(1-e^2)}\frac{1}{d_L}\sum_{n=1}^\infty \left\{A_n(e)\sin^2i\ \cos nl + \right. \nonumber \\ &\left.(1+\cos^2i)\left[\frac{S_n(e)-C_n(e)}{2}\cos[(nl+2\Phi)]+\right. \right . \nonumber \\
    & \left. \left. \frac{S_n(e)+C_n(e)}{2}\cos[(nl-2\Phi)]\right] \right\} \label{eq:strain plus}
\end{align}
which is identical to Eqs. \eqref{eq:gw_strain_hc} and \eqref{eq:gw_strain_hp} presented in the main text.

\section{Waveform systematics affecting recovery} \label{appendix:systematics}
\begin{figure}
    \centering
    \includegraphics[width=\linewidth]{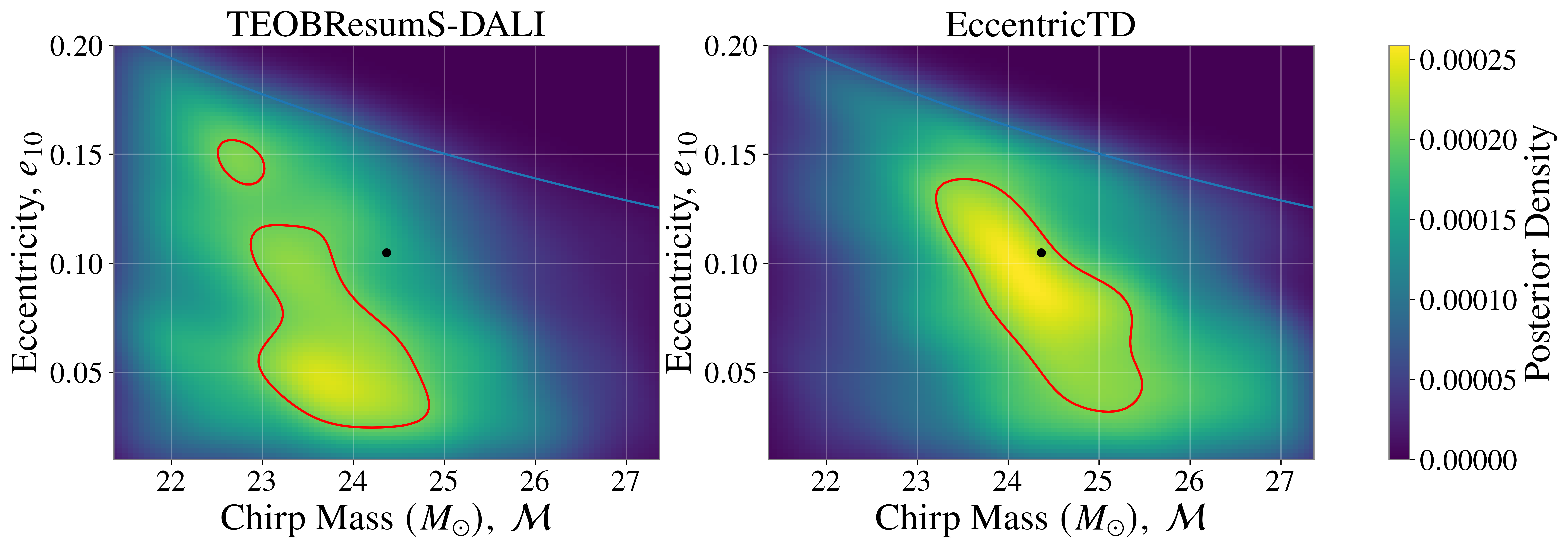}
    \caption{Comparing the recovery contours for identical zero spin systems but injected with TEOBResumS and EccentricTD. The injected parameters are overlaid onto the plot.}
    \label{fig:0spin_contours}
\end{figure}

EccentricTD is a time-domain inspiral-only model supporting non-spinning systems up to an eccentricity of 0.9. It utilises a Keplerian type parameterisation and modifies the existing TaylorT4 approximant to accommodate orbital eccentricities upto a 2PN order. It can accurately produce waveforms upto initial eccentricities of 0.9. On the other hand, TEOBResumS uses an \ac{eob} parameterisation to obtain an effective one body Hamiltonian that describes the two-body problem. It is able to support a wide variety of systems, includingnon-circularr, hyperbolic and non-planar orbits. It can maintain accuracy in the late inspiral via the use of Pade resummations and with a quasicircular NR informed ringdown model. These differences in modelling \acp{cbc} bleeds into the recovery of parameters using our approach. Given that the \ac{ecmm} was tuned with EccentricTD, we expect the best results when recovering systems injected with it. On the other hand, we find that attempting to recover parameters with TEOBResumS shows biases. This is clear from Fig \ref{fig:0spin_contours} where we compare the posterior densities when recovering signal injected with TEOBResumS and EccentricTD. We note that in addition to the shift in the overall distribution towards lower chrip masses, the posterior density is also weaker for the signal that was injected with TEOBResumS. This indicates that waveform systematics affect our ability to accurately recover the injected signal and improvements can be obtained by tuning the model to the most accurate waveform.

\section{Energy-Informed Pixel Extraction: Higher Harmonics} \label{appendix:ecmm}

\begin{figure}[hbt]
    \centering
    \includegraphics[scale=1]{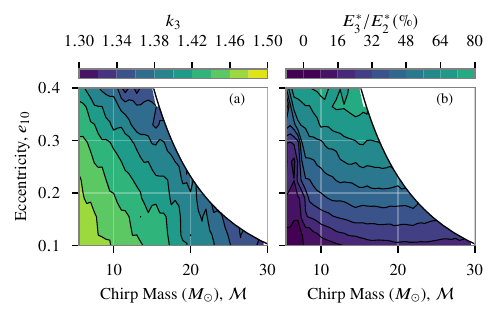}
    \caption{The left plot shows the contour of the scale factor corresponding to the first eccentric harmonic in the $\mathcal{M}-e_{10}$ space. The plot on the right shows the contour for the energy ratio between the fundamental and first harmonic track.}
    \label{fig:E3_scale}
\end{figure}

As discussed in Sec.~\ref{sec:EIEC}, we impose additional constraints while collecting the pixels along a frequency evolution track. In the previous work, a fixed number of nearest pixels were collected at each time slice, whereas in this work, we drop the pixels whose energy is greater than the energy of the pixel closest to the frequency track. 
In Fig.~\ref{fig:E3_scale}, we plot the contours of the scale factor of the first eccentric harmonic track in the $\mathcal{M}-e_{10}$ space along with the corresponding energy ratio between the fundamental and first harmonic. We find that the energy ratio iso-contours are roughly parallel to the $\mathcal{M}$ axis. Additionally, the variation along the $e_{10}$ axis is also very slow. These two factors contribute towards the broad contours during the recovery of the injected systems.


\bibliography{apssamp}
\end{document}